\documentclass[aip,apl,preprint]{revtex4-1}
\usepackage{graphicx,float,amssymb,subfigure,dcolumn,epstopdf}
\usepackage{multirow}
\usepackage{bm}
\begin{document}
 \title{Separating Inverse spin Hall voltage and spin rectification voltage by inverting spin injection direction}
 \author{Wenxu Zhang\footnote{Corresponding author. E-mail address: xwzhang@uestc.edu.cn}}
  \affiliation{State Key Laboratory of Electronic Thin Films and Integrated Devices,
  	University of Electronic Science and Technology of China, Chengdu, 610054, P. R. China}
 \author{ Bin Peng}
  \affiliation{State Key Laboratory of Electronic Thin Films and Integrated Devices,
  	University of Electronic Science and Technology of China, Chengdu, 610054, P. R. China}
 \author{Fangbin Han}
  \affiliation{State Key Laboratory of Electronic Thin Films and Integrated Devices,
  	University of Electronic Science and Technology of China, Chengdu, 610054, P. R. China}
 \author{Qiuru Wang}
  \affiliation{State Key Laboratory of Electronic Thin Films and Integrated Devices,
  	University of Electronic Science and Technology of China, Chengdu, 610054, P. R. China}
 \author {Wee Tee Soh}
 \affiliation{Center for Superconducting and Magnetic Materials, Department of Physics, National University of Singapore, 2 Science Drive 3, Singapore 117551, Singapore} 
  \author{Chong Kim Ong}
 \affiliation{Center for Superconducting and Magnetic Materials, Department of Physics, National University of Singapore, 2 Science Drive 3, Singapore 117551, Singapore} 
 \author{ Wanli Zhang}
 \affiliation{State Key Laboratory of Electronic Thin Films and Integrated Devices,
 University of Electronic Science and Technology of China, Chengdu, 610054, P. R. China}
 \date{\today}
\begin{abstract}
We develop a method for universally resolving the important issue of separating the inverse spin Hall effect (ISHE) from spin rectification effect (SRE) signal. This method is based on the consideration that the two effects depend on the spin injection direction: The  ISHE is an odd function of the spin injection direction while the SRE is independent on it. Thus, inversion of the spin injection direction changes the ISHE voltage signal, while SRE voltage remains. It applies generally to analyzing the different voltage contributions without fitting them to special line shapes.  This fast and simple method can be used in a wide frequency range, and has the flexibility of sample preparation.  
\end{abstract}
\maketitle

With the rapid development of spintronics, since the discovery \cite{fert, gruenberg}of GMR effects by Gr\"unberg and Fert in the late 1980's, manipulation of transportation and detection of spins are two of the central problems in this blooming science and technology. The inverse spin Hall effect (ISHE) is one of the effects experimentally demonstrated\cite{hirch} right after that,  where electric voltages are generated by pure spin nonequilibrium.  
\par In order to produce the spin current, pumping of  spins by microwave irradiation from the ferromagnetic(FM) materials to adjacent nonmagnetic(NM)  metal  materials were proposed\cite{ando, kurebayashi}. This is a breakthrough in this field, because the spins are effectively injected from FM to NM metals. DC voltages were generated in the NM metal due to the inverse spin Hall effect, which follows the line shape of the ferromagnetic resonance(FMR) spectra. Soon after, it was realized that in FM/NM bilayers, voltage at FMR has contributions not only from spin pumping, but also from spin rectification effect (SRE)\cite{hu11}. The voltages from SRE can not be neglected and suppressed except in special cases where the microwave electric field is kept to zero such as at the center of a microwave cavity with TE$_{011}$ mode\cite{saitoh}. However, it is difficult to explore the frequency characteristics of the FM/NM bilayers with microwave cavity because the cavity only works near its resonance frequency. It is convenient to study the ISHE at different frequencies with transmission line such as CPW \cite{hu11} or shorted microstrip\cite{wt}. However, the ISHE signal is in most cases contaminated by SRE because the SRE cannot be neglected in the transmission line and it may contribute voltages with the same line shape as the ISHE. Then, it is necessary to extract the ISHE signal from the mixed signal. The intricacies of separating the two effects have been solved by work of Hu\cite{ hu11} and Ong\cite{wt2}. It was shown that the two effects have endowed different dependences of the static magnetic field direction. Thus, rotation of the field in the film plane can be used to separate the two effects. However, due to the limitation  of linear response, the method is not applicable to high power cases.  Another generalized methods were proposed by  Hoffmann \emph{et al.}\cite{hoffmann13} where the different angular and field symmetries of the two effects were used to separate the two contributions. It does not rely on the linear approximation, and can be used in high power cases to study the nonlinear effects.
\par In this \emph{Letter}, we proposed another universal method to separate the SRE and ISHE voltage by simplifying the measurement to two steps.  Considering that the mixed contributions consist of odd and even function with respect to the spin injection direction, we separate them by taking two measurements where the spin injection are inverted. It reveals that the SRE has both Lorentian and dispersive contributions to the voltage while the ISHE has only a Lorentzian contribution. These voltages are also a function of the microwave frequency, but their ratio remains almost constant. 
\par We begin by pointing out the symmetry properties of the photovoltages. As shown in the previous work\cite{hu11},  the anisotropic magnetoresistance contributes to the DC voltage because of the phase differences of the microwave current ($\vec{j}$) and magnetization ($\vec{m}$) precession, which roots on the broken rotational invariance of FM as in the two-band model\cite{two-band} for spin transport. 
\par
\begin{figure}
	\includegraphics[width=0.48\textwidth]{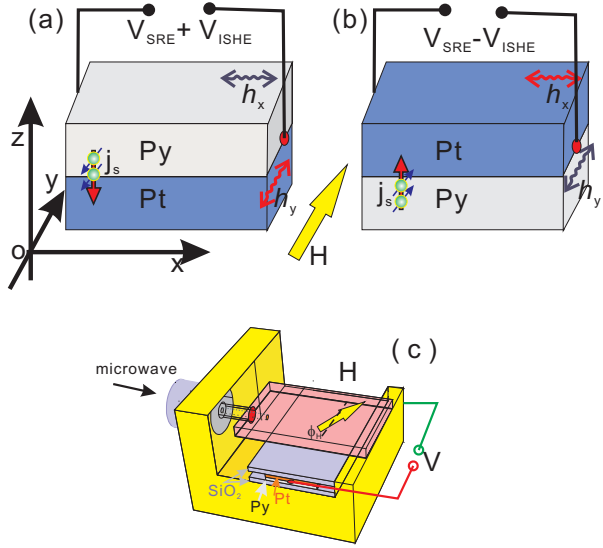}
	\caption{Sketch for the dc voltage induced in FM/NM bilayer by (a) normal positioned samples (b) flipped samples and (c) our shorted microstrip fixture.}\label{fig:sketch}
\end{figure}

 The voltage V$_{SRE}$ can be expressed by 
\begin{equation}
V_{SRE}\propto\langle(\vec{m}\cdot\vec{e}_x)j_x\rangle \vec{e}_x\cdot\vec{e}_H.\label{equ:sr}
\end{equation}
where $\vec{m}=\vec{M}(t)-\vec{M}$ is the magnetization pumped by the FMR, which propagates in the FM and from FM to NM layer. $\vec{e}_{x,H}$ are the directions of the $x$-axis in our coordinate and the static magnetic field (H) as shown in Fig. \ref{fig:sketch}.  
The diffusion of the spins from Py to NM layer gives rise to a DC voltage due to the spin-orbit coupling. The voltage (V$_{ISHE}$) is 
expressed by 
\begin{equation}
V_{ISHE}\propto(|\vec{m}\cdot\vec{e}_H|\omega_r)\vec{e}_x\cdot(\vec{e}_z\times\vec{e}_H).\label{equ:ish}
\end{equation}
where $\omega_r$ is the FMR frequency involved in both $V_{SRE}$ and $V_{ISHE}$. 
\par The total photon voltage is expressed as the summation of both: $V_{Ph}=V_{SRE}+V_{ISHE}$.  As clearly shown in Equ.(\ref{equ:sr}) and (\ref{equ:ish}), The $V_{ISHE}$ is an odd function of $\vec e_z$ while $V_{SRE}$ is independent on it.
This is quite understandable because $\vec e_z$ is related to the spin diffusion direction, which is irrelevant to produce $V_{SRE}$. Thus, when only the spin diffusion direction is reversed  as in Fig. \ref{fig:sketch} (b) in our coordinate, the new voltage is   $V_{Ph}^{I}=V_{SRE}-V_{ISHE}$. In this case, the two contributions can be separated by summation and subtraction of the two measurement: $V_{SRE}=\frac{1}{2}(V_{Ph}+V_{Ph}^{I})$ and  $V_{ISHE}=\frac{1}{2}(V_{Ph}-V_{Ph}^{I})$.  This symmetry property was demonstrated by measuring the voltage in samples with reversed stacking order of NM/CoFeB on thermally oxidized Si substrate\cite{kim}. However, because film quality may be quite sensitive to the underlayers, cautions should be taken when extracting the informations.  We thus propose the following measurement schemes.
\par  The measurement was done by our shorted microstrip fixture \cite{han} which can work up to 8 GHz as schematically shown in Fig.\ref{fig:sketch}(c). Our samples include a permalloy(Py, Ni$_{80}$Fe$_{20}$ )(20 nm)/SiO$_2$(0.2 mm substrate) monolayer and Pt(10 nm)/Py(20 nm)/SiO$_2$(0.2 mm substrate) bilayer. Both samples have lateral dimensions of 5 mm $\times$ 10 mm. The monolayer serves as a control where the measured voltage should be the same when the sample is flipped. In order to put the samples at the same positions in the fixture and minimize the differences of to the microwave field before and after sample flipping, we covered the samples with the same material with the same dimensions as the substrate. We obtained the voltage by lock-in techniques (SR830, Standford research system) with microwave source power provided by Rohde \& Schwarz(SMB 100 A). At the fixed microwave frequency, we sweep the static magnetic field so that FMR was observed.
\begin{figure}
	\includegraphics[width=0.48\textwidth]{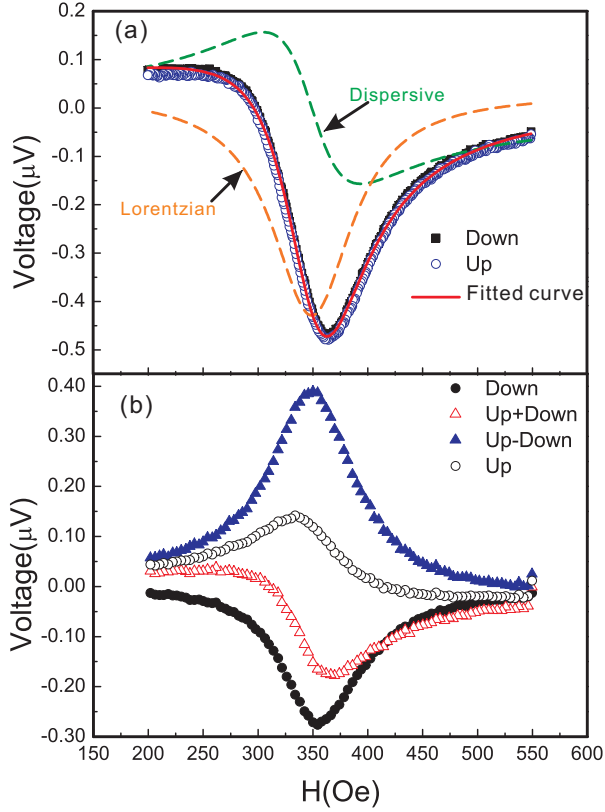}
	\caption{\label{fig:ishe1}Voltages measured at different applied field in (a) Py/SiO$_2$ and (b) Py/Pt/SiO$_2$ in the two different configurations. The measurement is conducted at 5.2 GHz. ``Up'' and ``Down'' mean sample with film upwards and downwards, respectively. }
\end{figure}
\par In a ferromagnetic monolayer, only SRE was generated. The voltage measured in the film is shown by hollow circles in Fig.\ref{fig:ishe1}(a). The curves of the monolayer shows clearly a combination of the Lorentzian and dispersive contributions. The relative strength of the two contributions is a function of the phase angle of the microwave electric and current fields as pointed out by  Harder \emph{et al.}\cite{hu11}. When the samples are flipped, the signal is identical to the previous one as shown by filled circles in Fig.\ref{fig:ishe1}(a), which is expected by the symmetry of the SRE voltage. The signal of the bilayer under the normal (``Up'') and flipped (``Down'') configuration are shown in Fig.\ref{fig:ishe1}(b). A clear difference  of the two curves comes from the inversed spin diffusion direction in our coordinate. The contributions from the ISHE and SRE can be obtained by simple  subtraction (``Up-Down'') and summation (``Up+Down'') of the two curves, respectively. A factor of 0.5 were taken into account when the data were calculated. Clearly, the ISHE voltage is Lorentzian type while the SRE is a combination of the two types. The SRE voltages  from the monolayer and the bilayer are proportional to each other  because of the differences of the resistance of the samples: the Pt layer on the Py thin film acts as  an electrical shunt.  
\par When we compare the voltages obtained by our fixture with those measured by CPW, the latter  usually being of  the order of tens of $\mu V$, the voltage measured by the former is about one order smaller. This is due to the microwave magnetic field being smaller in our setups. The samples are about 0.5 mm away from the conducting strip. However, the signal is clearly above the noise level of the lock-in amplifier. The advantage of our methods is that samples and the microstrips are reusable, so that comparisons of different samples and other material characterizations of the samples can be readily done, while in the CPW setups, the magnetic films are deposited directly on the CPW substrate.   
\begin{figure}
    \includegraphics[width=0.48\textwidth]{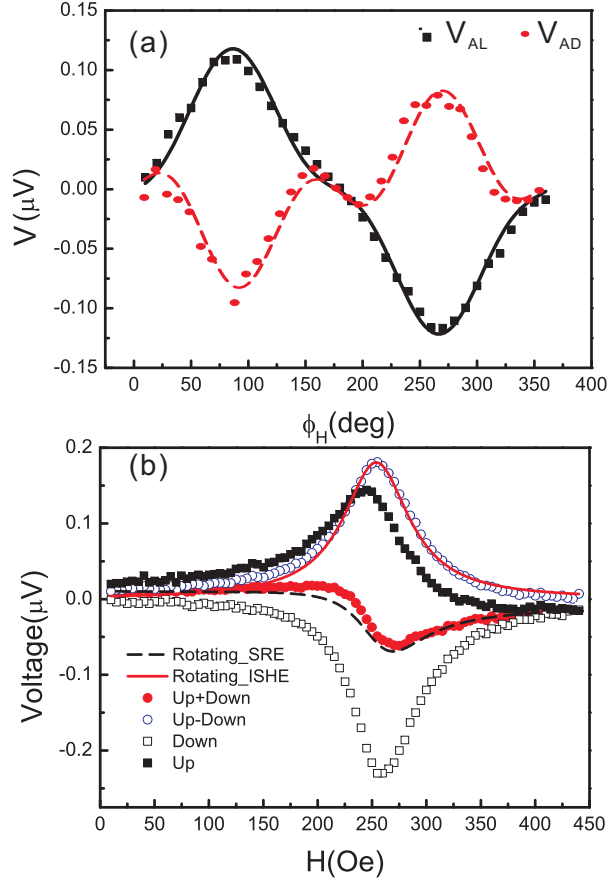}
	\caption{\label{fig:angular}(Color online)Magnetic field angular ($\phi_H$) dependent Lorentzian (V$_{AL}$)and dispersive (V$_{AD}$) voltage amplitudes of Pt/Py/SiO$_2$ (a) and the same sample measured by the new method (b). The $\phi_H$ dependent lines in (a) are fitted to the theoretical formulas. All the measurements were done at 4.4 GHz.}
\end{figure}
\par In order to have a comparison of our measurement to those of rotation samples as used by Ong\cite{wt2} and repeated by us\cite{han}, we show the separated voltages of the same sample measured at different angles in Fig.\ref{fig:angular} (a) by the same fixture. The data were rendered from sets of measurement at different magnetic field angles $\phi_H$. At each of these fixed $\phi_H$, a sweeping of the magnetic field was done. Then, their voltages under different fields were fitted to the Lorentzian and dispersive line shape. The amplitudes V$_{AL}$ and V$_{AD}$ were shown in Fig.\ref{fig:angular}(a) by filled squares and circles. The voltage contributions from the ISHE and the SRE were obtained by fitting the curves according to Equ.(\ref{equ:val}) and (\ref{equ:vad})\cite{wt2,han}.
\begin{widetext}
\begin{eqnarray}
V_{AL}&=&-\sin\Phi[V_{MRz}\sin\phi_H\cos(2\phi_H)-V_{MRx}\sin\phi_H\sin(2\phi_H)]-
V_{AHE}\cos\Phi\sin\phi_H+V_{ISHE}\sin^3\phi_H\label{equ:val}\\
V_{AD}&=&\cos\Phi[V_{MRz}\sin\phi_H\cos(2\phi_H)-V_{MRx}\sin\phi_H\sin(2\phi_H)]- 
V_{AHE}\sin\Phi\sin\phi_H\label{equ:vad}
\end{eqnarray} 
\end{widetext}
 \par  We obtain the ``Up'' and ``Down'' curves at a specified $\phi_H$(=90$^\circ$), where the ISHE and SRE take their maxima.  The curves are shown in Fig.\ref{fig:angular} (b). The red solid line is reproduced from the V$_{ISHE}$ data obtained by the rotation sample method above. It follows well the curve (hollow blue circles) obtained by our new method.  The V$_{ISHE}$ peak values obtained is 0.180 $\mu V$, which is comparable with 0.182 $\mu V$ obtained above. The differences is within few percent of the voltage. The SRE curves obtained by the two methods are shown by the black dashed line and red filled circles. The small deviation below the resonant field may come from the uncertainty of the data fitting.  Thus, our methods provide the same information as the rotation method but with much reduced number of measurements.

\begin{figure}
	\includegraphics[width=0.48\textwidth]{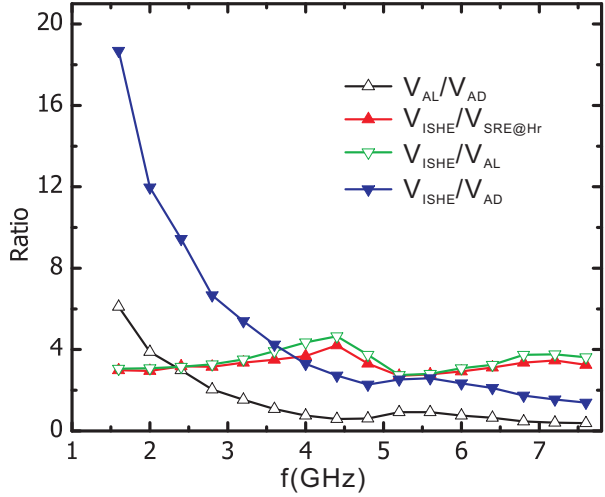}
	\caption{\label{fig:v-freq} (Color online)The ratio of ISHE to the different parts of SRE voltages  measured at different frequencies.}
\end{figure}      
\par As pointed by Hu \emph{et al.} \cite{hu11}, when there are contributions of Lorentzian and dispersive types to the SRE, cares must be taken to get reasonable results.  The phase difference of the microwave electric field and dynamic magnetic moment may change with microwave frequency.  As argued in the work of Hoffman \emph{et al}\cite{hoffmann13}, since equation (1) and (2) predicts, V$_{ISHE}$$\propto$V$_{SRE}$$\propto$ $P/\omega_r$, we may thus use this relation to check for consistency in our obtained results over the measured frequency range.  As can be seen in Fig.\ref{fig:v-freq}, the ratio of V$_{AL}$ to V$_{AD}$, which is dependent on the phase, changes in a wide range of the frequencies and shoots up in the low frequency range. If we just took one of them, say V$_{AD}$, the ratio of V$_{ISHE}$ to V$_{AD}$ shoots up at low frequency. In our measurements,  the main contribution is from the V$_{AL}$, because it is an ``h$_x'$'' dominated FMR as classified in the work \cite{hu11}.  Thus,  we may take the ratio of V$_{ISHE}$ to V$_{AL}$ separated from V$_{SRE}$, or that of V$_{ISHE}$ to V$_{SRE}$ at the resonance field, since the term with V$_{AD}$ diminishes at resonance although V$_{AD}$ itself is finite. The results are shown in Fig. \ref{fig:v-freq}.  As expected, the ratios of V$_{ISHE}$ to V$_{AL}$ and to V$_{SRE\\@Hr}$ are almost constant in the whole frequencies. However, in this wide frequency range there is a strong phase mixing between the $e$ and $h$ fields, especially when the frequency is lower than 4 GHz, as reflected by the variation of V$_{AL}$/V$_{AD}$.     

\par In summary, we have proposed a method to separate  the ISHE and SRE voltages in the sample by flipping the samples inside a shorted 
microstrip fixture. The proposal is based on the fact that ISHE is an odd function of spin injection direction while SRE is not relevant to it. This method can also be generalized to other cases, like when the spin Seebeck effect is involved, where the voltage has a different coordinate parity with respect to SRE. Since the separation is independent of assumption of linear response of the magnetization to the microwave field, our methods  is not limited by the microwave frequency and power.    
\newline
\par Financial support from NSFC(61471095), ``863''-projects (2015AA03130102) and Research Grant of Chinese Central Universities (ZYGX2013Z001) are acknowledged.

\end{document}